\documentclass[aps,pre,reprint,superscriptaddress,nobibnotes,nofootinbib,floatfix]{revtex4-2}

\usepackage[T1]{fontenc}
\usepackage{lmodern}
\usepackage{amsmath,amssymb,graphicx}
\usepackage[hidelinks]{hyperref}
\usepackage{url}
\DeclareUnicodeCharacter{03BC}{\ensuremath{\mu}}
\DeclareUnicodeCharacter{00B5}{\textup{\textmu}}
\DeclareUnicodeCharacter{00D7}{\ensuremath{\times}}
\DeclareUnicodeCharacter{2212}{\ensuremath{-}}
\DeclareUnicodeCharacter{00B2}{\textsuperscript{2}}
\DeclareUnicodeCharacter{00B3}{\textsuperscript{3}}
\DeclareUnicodeCharacter{2080}{\textsubscript{0}}
\DeclareUnicodeCharacter{2081}{\textsubscript{1}}
\DeclareUnicodeCharacter{2082}{\textsubscript{2}}
\DeclareUnicodeCharacter{2083}{\textsubscript{3}}
\DeclareUnicodeCharacter{2084}{\textsubscript{4}}

\providecommand{\pandocbounded}[1]{\begingroup
  \sbox{0}{#1}%
  \ifdim\wd0>\linewidth \resizebox{\linewidth}{!}{\usebox{0}}%
  \else \usebox{0}\fi
\endgroup}

\begin{document}

\title{Measuring the Roughness of Quasi-Circular Interfaces at Finite Resolution}

\author{Ralph Bulanadi}

\email{bulanadira@ornl.gov}

\affiliation{Center for Nanophase Materials Sciences, Oak Ridge National Laboratory, Oak Ridge, TN 37831, USA}

\author{Asraful Haque}

\affiliation{Center for Nanophase Materials Sciences, Oak Ridge National Laboratory, Oak Ridge, TN 37831, USA}

\author{Sumner B. Harris}

\affiliation{Center for Nanophase Materials Sciences, Oak Ridge National Laboratory, Oak Ridge, TN 37831, USA}

\author{Rama Vasudevan}

\affiliation{Center for Nanophase Materials Sciences, Oak Ridge National Laboratory, Oak Ridge, TN 37831, USA}

\author{Yongtao Liu}

\email{liuy3@ornl.gov}

\affiliation{Center for Nanophase Materials Sciences, Oak Ridge National Laboratory, Oak Ridge, TN 37831, USA}

\begin{abstract}

Interface roughness characterizes the boundary fluctuations across
different length scales, which can reveal information about a material's
pinning landscape and collective interfacial behaviour. Finite sampling
and image resolution can bias the measured scaling properties, however.
These effects can particularly complicate comparisons between circular
domains of different sizes, where the number of resolved boundary
points, and the relationship between angular and physical separations,
vary with the interface radius. Here, we benchmark the recovery of the
roughness exponent and amplitude using synthetic self-affine interfaces
generated by Fourier synthesis, mapped onto circular contours, and
rasterized onto a pixel grid. We first compare displacement-correlation
(\(B(r)\)) and structure-factor (\(S(q)\)) estimates for different
interface lengths. To address non-uniform angular and radial spacing in
a circular interface, we assess a reconstruction procedure that weights
slope minimization against medial interface position. We find that while
spectral estimates were most effective in the original interfaces, this
advantage did not persist after circular mapping. Moderate smoothing was
found to improve parameter recovery, but stronger smoothing can bias
both the scaling exponent and roughness amplitude. These results,
verified on both synthetic datasets, and an experimental ferroelectric
system, identify limitations of circular-interface roughness
measurements and provide a basis for further investigations into the
evolution of circular domains under finite resolution experiments.

\end{abstract}

\maketitle

\begingroup
\renewcommand{\thefootnote}{}

\footnotetext{Notice: This manuscript has been authored by UT-Battelle,
LLC, under Contract No. DE-AC0500OR22725 with the U.S. Department of
Energy. The United States Government retains and the publisher, by
accepting the article for publication, acknowledges that the United
States Government retains a non-exclusive, paid-up, irrevocable,
world-wide license to publish or reproduce the published form of this
manuscript, or allow others to do so, for the United States Government
purposes. The Department of Energy will provide public access to these
results of federally sponsored research in accordance with the DOE
Public Access Plan
(\url{http://energy.gov/downloads/doe-public-access-plan}).}

\endgroup


Interface roughness provides a statistical description of boundary
fluctuations and their dependence on observation scale. In ferroic
materials, roughness measurements connect the geometry of domain walls
to the disorder and elastic interactions that govern their
behavior~\cite{ref1,ref2,ref3,ref4}, providing information about a
material's pinning landscape and collective interfacial behavior.

The reliability of roughness estimators has been examined extensively
using synthetic profiles with prescribed scaling
properties~\cite{ref5,ref6}. The accuracy of a recovered exponent is
known to depend on the estimator, profile length, and underlying
fluctuations, as well as show substantial sample-to-sample
variation~\cite{ref7}. Finite spatial resolution introduces additional
limitations: a finite probe size can generate spurious small-scale
behavior~\cite{ref8}; background removal and other processing operations
can introduce systematic bias~\cite{ref9}; and fluctuations below the
measurement floor may be unresolved. As such, agreement with a power-law
over a selected interval does not necessarily establish that the fitted
parameters accurately describe the underlying interface.

Closed, quasi-circular domains, which can be found in bubble-like
ferromagnetic or ferroelectric domains are of particular interest for
low-energy memory applications~\cite{ref10,ref11,ref12,ref13}.
Correlation-based roughness analysis has recently been applied to
bubble-like ferromagnetic domains to establish self-affine
behavior~\cite{ref14,ref15}. However, studies of nanoscale ferroelectric
domains~\cite{ref11}, where domains can only be resolved with relatively
few pixel sizes, may require additional care. Boundary points extracted
from a Cartesian grid are not uniformly spaced in angle, and as such
both their radial and angular uncertainties vary around the contour.
Rasterization and boundary extraction can therefore introduce apparent
fluctuations even for a perfectly circular interface.

These issues become increasingly prominent when studying expanding or
contracting domains, where the number of resolved boundary points may
change substantially throughout an experiment. Changes in the fitted
exponent or amplitude may thus reflect changes in sampling conditions or
changes in reconstruction bias, in addition to true changes in the
roughness itself. The increasing adoption of automated and autonomous
microscopy also allow for larger datasets and systematic exploration of
domain behavior~\cite{ref16,ref17,ref18,ref19,ref20}, but are still
limited by the resolution of the experiment; data processing methods
that allow recovery of these parameters in experiments with finite
resolution is thus of key importance. Quantifying these effects and
benchmarking the accuracy of roughness recovery in interfaces through
the circular-mapping and rasterizing process can therefore support more
conclusive studies on domain growth and pinning.

\begin{figure*}[htb]
\centering

\includegraphics[width=\linewidth,height=.72\textheight,keepaspectratio]{\detokenize{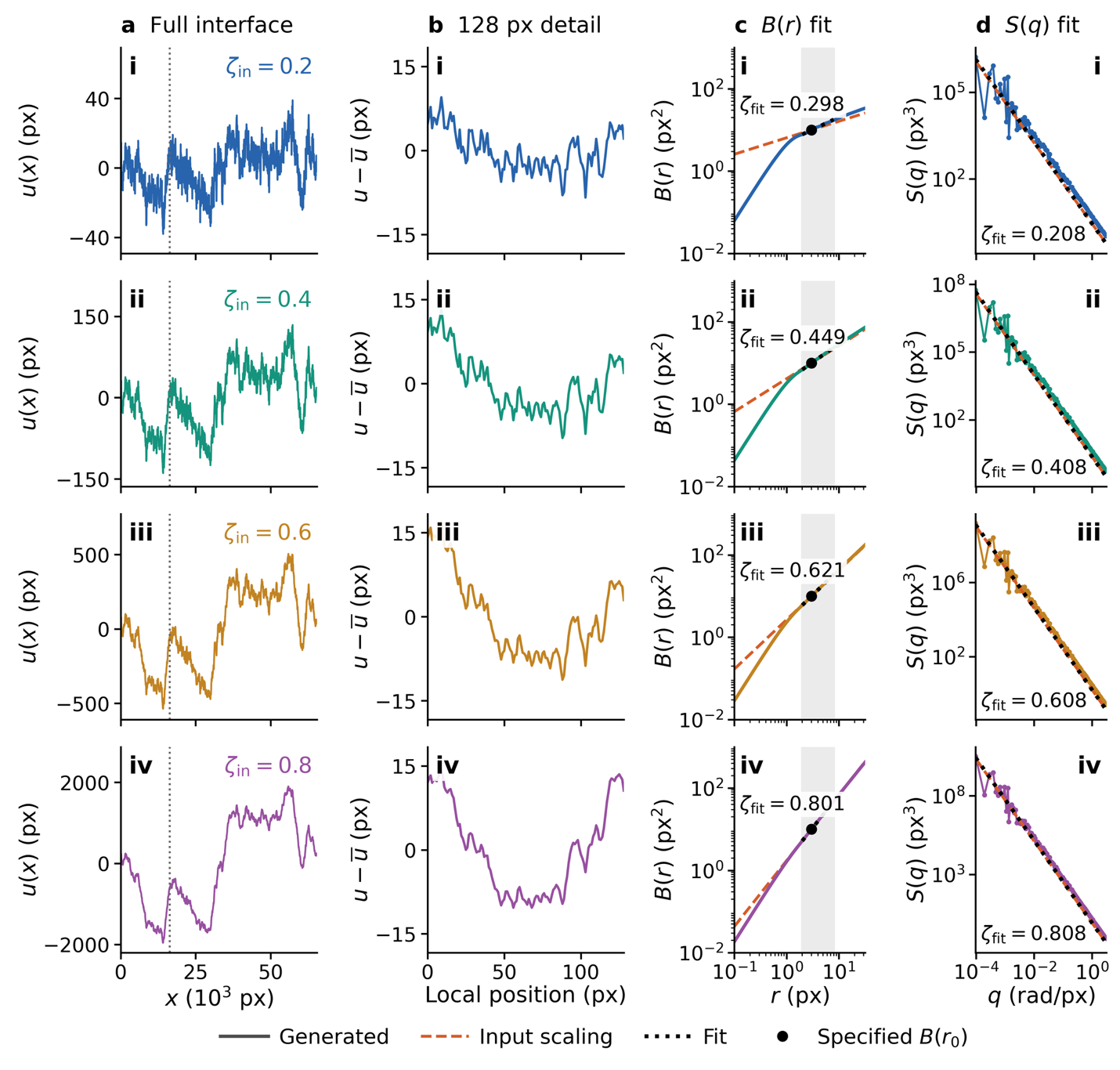}}

\caption{Synthetic periodic interfaces and their real-space and
reciprocal-space roughness measures. (a) shows the full interface, (b) a
mean-subtracted 128-pixel segment, (c) the displacement-difference
correlation function exponents \(B(r)\), and (d) the structure factor
\(S(q)\). (i-iv). correspond to prescribed roughness exponents
\(\zeta = 0.2,\ 0.4,\ 0.6,\ 0.8\) respectively. For (c,d), dashed orange
lines indicate the prescribed scaling, and dotted black lines show
power-law fits; gray shading marks the fitting interval for \(B(r)\).
Black markers identify the specified value \(B_{0}\). Insets report the
prescribed and fitted exponents.}

\label{fig:1}

\end{figure*}

In this work, we benchmark the recovery of roughness parameters from
synthetic and experimental interfaces generated by Fourier synthesis
with prescribed scaling parameters. We first examine finite-size effects
in their original profiles, then examine the impact of circular mapping,
rasterization, and boundary reconstruction. We compare
displacement-correlation (\(B(r)\)) and structure-factor (\(S(q)\))
estimates to evaluate how these factors affect reconstruction. In doing
so, we identify conditions to improve agreement between prescribed and
measured interfaces to provide a basis for further study into evolving
circular interfaces.

For a self-affine interface, the interface roughness can be quantified
by the displacement-difference correlation function~\cite{ref4,ref7}:

\begin{figure*}[htb]
\centering

\includegraphics[width=\linewidth,height=.72\textheight,keepaspectratio]{\detokenize{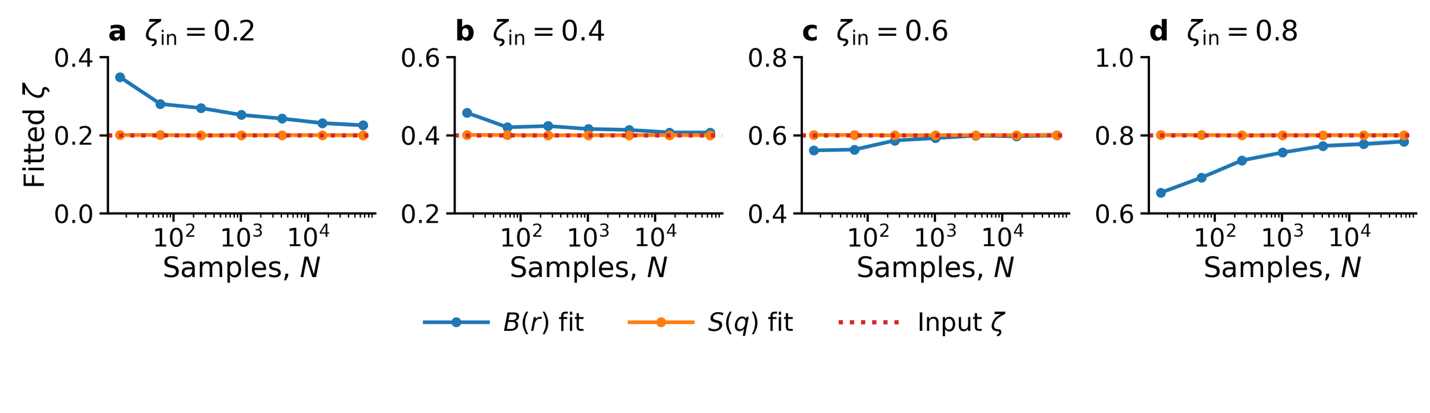}}

\caption{Finite-size dependence of roughness-exponent recovery for
periodic interfaces. (a--d) correspond to prescribed roughness exponents
\(\zeta = 0.2,\ 0.4,\ 0.6,\ 0.8\) respectively. Blue and orange curves
show exponents obtained from fits to \(B(r)\), and \(S(q)\),
respectively, as a function of the number of points per interface,
\(N\). Ensembles target a total of \(2^{17}\) sampled points at each
interface length, averaged over 20 independent repetitions. Red dotted
lines indicate the prescribed exponents.}

\label{fig:2}

\end{figure*}

\begin{equation}
B(r) = \left\langle \left\lbrack u(x + r) - u(x) \right\rbrack^{2} \right\rangle
\label{eq:1}
\end{equation}where \(u(x)\) is the displacement of the domain wall from a reference
contour, \(r\) is the distance along the interface, and the angular
brackets denote averaging over boundary positions. For self-affine
interfaces, this function fits to power-law scaling:

\begin{equation}
B(r) = B_{0}\left( \frac{r}{r_{0}} \right)^{2\zeta}
\label{eq:2}
\end{equation}
where \(\zeta\) is the roughness exponent, which describes how
displacement differences grow with separation, and \(B_{0}\) describes
the mean-squared magnitude at the reference length scale \(r_{0}\).

To benchmark the recovery of these roughness parameters, we generate
synthetic periodic interfaces by Fourier synthesis:

\begin{equation}
u(z) = \sum_{k}^{}\left\lbrack a_{k}\cos\left( q_{k}z \right) + b_{k}\sin\left( q_{k}z \right) \right\rbrack
\label{eq:3}
\end{equation}

\begin{equation}
q_{k} = \frac{2\pi k}{L}
\label{eq:4}
\end{equation}
where L is the interface length, and the coefficients are independent
variables with variances:

\begin{equation}
\text{Var}\left( a_{k} \right) = \text{Var}\left( b_{k} \right) = V_{k} = Ck^{- (1 + 2\zeta)}
\label{eq:5}
\end{equation}
similar to as produced by Ref.~\cite{ref21}. These interfaces can
alternatively be characterized in reciprocal space using the structure
factor:

\begin{equation}
S(q) = \frac{1}{L}\left\langle \left| \int_{0}^{L}{u(x)e^{- iqx}dx} \right|^{2} \right\rangle
\label{eq:6}
\end{equation}
which measures the contribution of fluctuations at wavelength
\(\lambda = 2\pi/q\). The generated interfaces would here be expected to
follow:

\begin{equation}
S\left( q_{k} \right) = \frac{L}{2}V_{k} \propto q_{k}^{- (1 + 2\zeta)}
\label{eq:7}
\end{equation}

The scaling exponent can therefore be obtained from the log--log slopes
of either \(B(r)\) or \(S(q)\). While \(B(r)\) directly expresses the
displacement differences at physically relevant distances, the fitted
exponent can depend strongly on the selected distance range: Shorter
interfaces restrict the size of the largest possible wavelength, and
resolution meanwhile restricts the smallest possible wavelength.
Spectral analysis measures the fluctuations by wavelength and can more
directly probe the scale-dependent changes, but requires more extensive
statistics to reduce error and fluctuations. We therefore benchmark both
measures on the same synthetic interfaces: first in their original
periodic form; after circular mapping; then finally after reconstruction
on a rasterized grid. This allows separation of the baseline effects of
finite interface length from any additional errors introduced by
finite-resolution measurements.

Some example rough interfaces, with the input \(\zeta\), which we denote
as \(\zeta^{\text{in}}\), ranging from 0.2 to 0.8, and each interface including
65536 points with \(r_{0} = 3\), \(\ B_{0} = 10\), is shown in Fig.~\ref{fig:1}.
Figs.~\ref{fig:1}a and \ref{fig:1}b show the interface at different resolutions; interfaces
with a higher \(\zeta\) show more correlated noise, with larger absolute
displacements over larger distances. Fig.~\ref{fig:1}c shows the fit for \(B(r)\),
which shows an imperfect fit; the fitting region has an impact on the
extracted scaling parameter, and, particularly for lower values of
\(\zeta^{\text{in}}\), the measured value of \(\zeta\) is quite distinct from
\(\zeta^{\text{in}}\). In contrast, the values extracted for \(S(q)\) shows
stronger agreement that is less dependent on the fitting region.


\begin{figure*}[htb]
\centering

\includegraphics[width=\linewidth,height=.72\textheight,keepaspectratio]{\detokenize{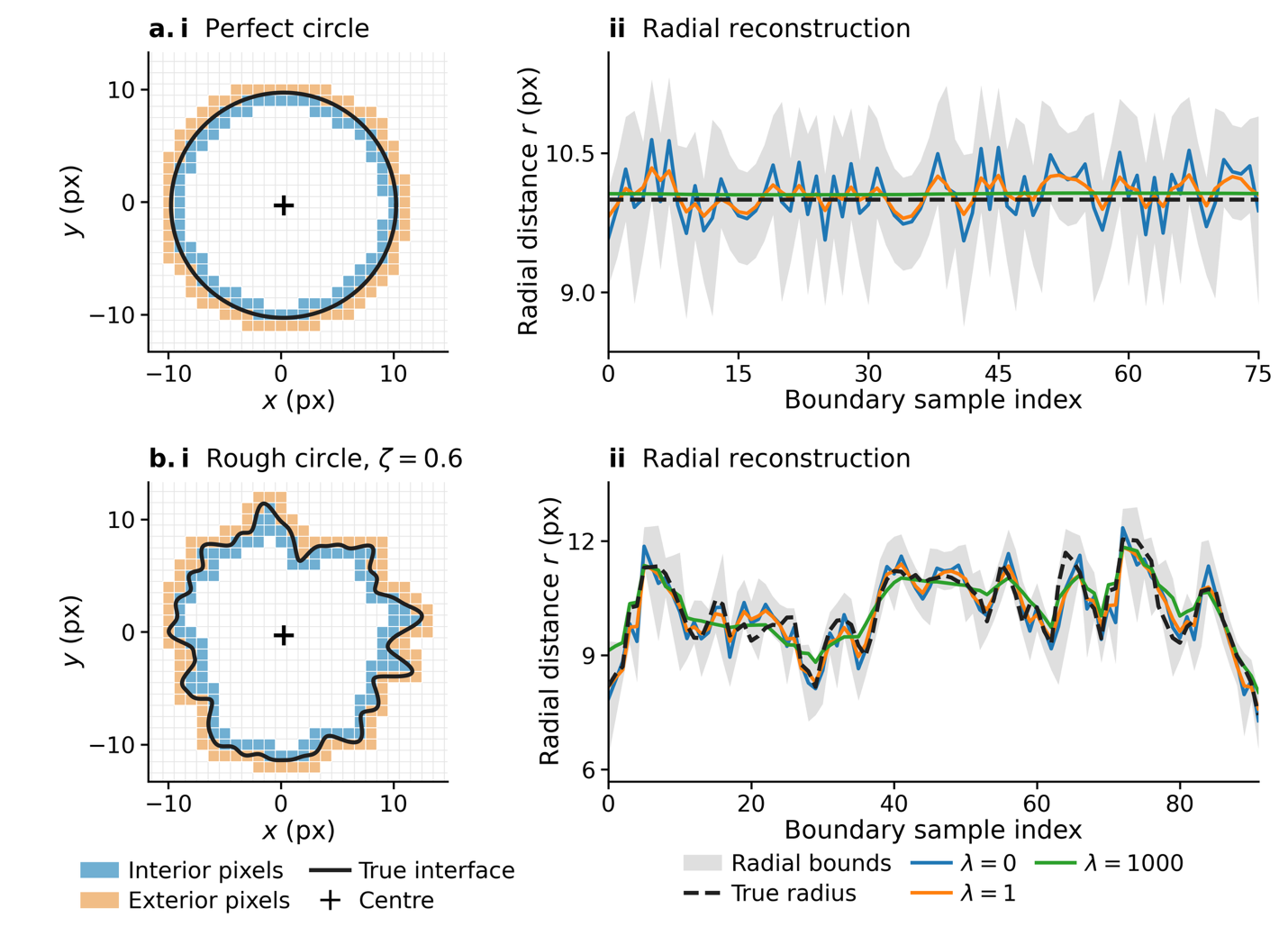}}

\caption{Rasterization and constrained reconstruction of circular
boundaries. (a) A perfect circle of radius 10 px and (b) a rough
circular interface with prescribed exponent \(\zeta = 0.6\). (i) Shows
the true contours (black dashed lines), interior boundary pixels (filled
blue squares), dilated exterior pixels (filled orange squares), and
reference centers (black crosses). (ii) Shows radial positions along the
extracted boundaries. Gray shading indicates the radial bounds obtained
from the inner and outer pixel centers. Colored curves show
reconstructions with smoothing strengths \(\lambda = 0,\ 1,\ 1000\),
while black dashed curves show the true radial profiles.}

\label{fig:3}

\end{figure*}

These differences highlight an initial problem with measuring the
scaling parameter in finite, real datasets: the accuracy of the
measurement \(\zeta\) from \(B(r)\) is dependent on the interface
length. This is highlighted in Fig.~\ref{fig:2}. Here, interfaces were generated
with \(N\) samples per interface ranging from \(2^{4}\) to \(2^{16}\).
Multiple interfaces were generated such that \(2^{17}\) samples were
generated for each interface length. The scaling parameter was then
measured (for measurements of \(B(r)\), the fit window was taken in the
interval between \(N^{1/2}\) and \(N^{1/6}\)). Here, the values of
\(\zeta\) gathered from the \(B(r)\) fit differ by as much as 0.15 with
\(N \approx 10\), even with extensive repetition. These approach
\(\zeta^{in}\ \)at much higher values of \(N\). The scaling parameter
measured from \(S(q)\) remains accurate over all conditions. This was
found to be highly reproducible: this process was repeated 20 times,
with their average values plotted in Fig.~\ref{fig:2}. All repetitions fell within
3\% of their mean value.

The sampling sensitivity of the scaling parameter \(\zeta\) when
measured from \(B(r)\) is of immense importance for the measurement of
mobile circular interfaces, where, as a boundary grows or expands,
resolution and sampling sensitivity may increase. A measurement of
\(\zeta\) from \(B(r)\) in a growing interface may, for example,
transition towards higher or lower values, but this may only be a
measurement of a growing sample size. Further, the maximum size of
\(B_{0}\) is more bounded in smaller domains than they are in larger
domains.

\begin{figure*}[htb]
\centering

\includegraphics[width=\linewidth,height=.72\textheight,keepaspectratio]{\detokenize{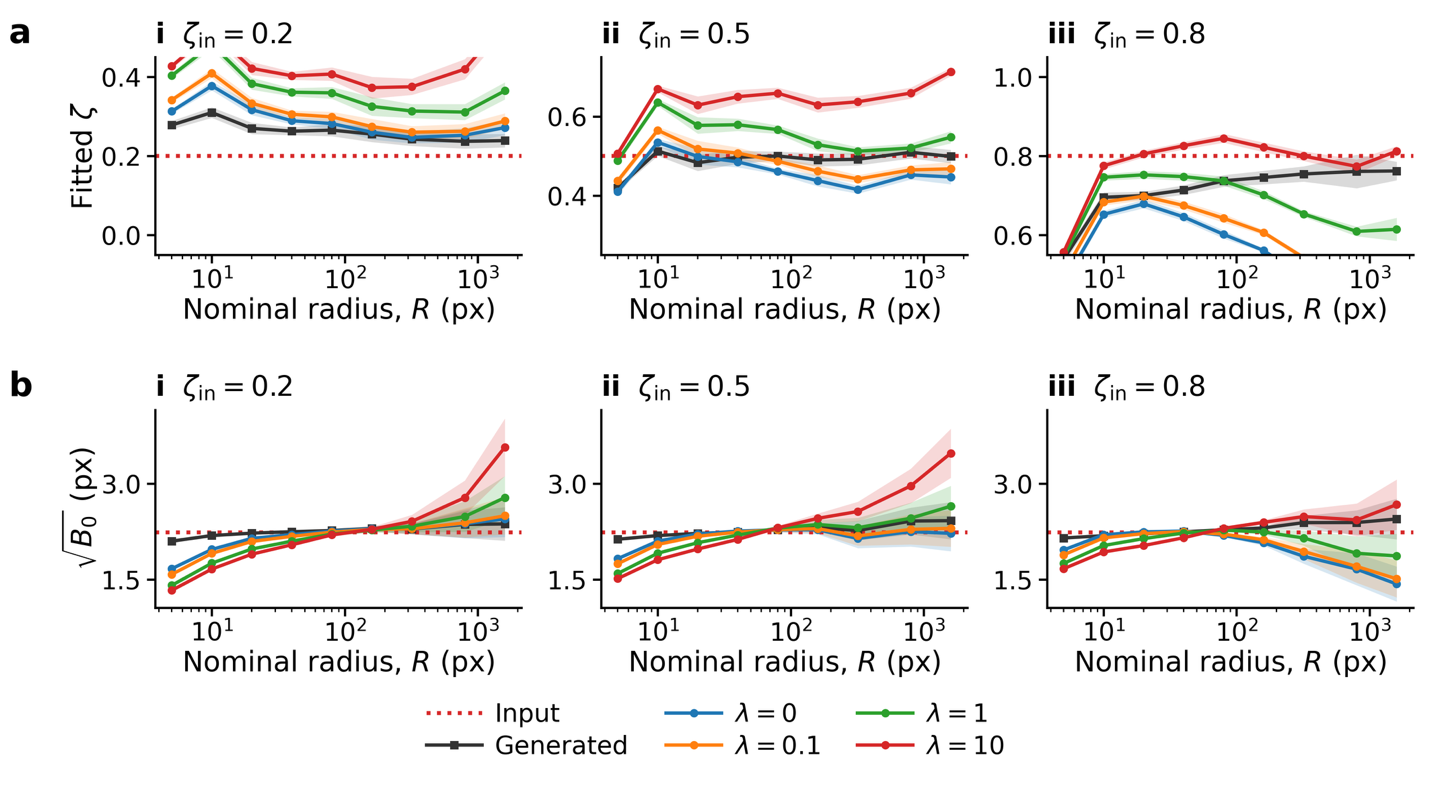}}

\caption{Recovery of circular-interface roughness at a fixed angular
reference separation \(r_{0} = \frac{\pi}{10}\,\mathrm{rad}\). (a) Shows
the exponent \(\zeta\) obtained from displacement-correlation fits; (b)
shows the corresponding amplitude \(\sqrt{B_{0}}\). (i--iii) correspond
to prescribed exponents \(\zeta = 0.2,\ 0.5,\ 0.8\), with
\(B_{0} = 5\,\mathrm{px}^{2}\). Results are plotted against nominal
radius \(R\). Red dotted lines indicate the prescribed values, and solid
black curves show estimates from the generated interfaces before
rasterization. The remaining curves show estimates after rasterization
and reconstruction with \(\lambda = 0,\ 0.1,\ 1,\ 10\).}

\label{fig:4}

\end{figure*}

Beyond this changing sample size, measurements of finite linear
boundaries have other key differences to measurements of finite circular
boundaries. Pixels that are regularly spaced in \((x,y)\) coordinates
are not regularly spaced in \((r,\theta)\) coordinates, which means that
both our spacing along the boundary, distances across the boundary, and
all associated errors are no longer uniform. An example of this is shown
in Fig.~\ref{fig:3}ai. where a circle with a known radius is constructed on a
grid. The interface always exists between the centers of the inner pixel
boundary, and the centers of the exterior pixel boundary. The error in a
position, even if the average of these two boundaries is taken, expected
to be as much as \(\pm \sqrt{2}\), creating an artificial roughness
value. This is shown in Fig.~\ref{fig:3}aii: while the true interface (black
dashed line) is entirely within the bounds (grey shaded area), taking
the middle of these bounds (solid blue line) does not accurately
represent the true interface.


An alternative formulation could be to fit a line between the minimum
and maximum bounds of each pixel that minimizes the slope. Specifically,
by minimizing the penalty:
\(J = \oint_{}^{}\left( \frac{d\rho}{d\theta} \right)^{2}d\theta\) where
\(\rho\) is the radial distance from the fitted center \(P\). This can
create the opposite problem, however, where real variations are
minimized. We therefore introduce a mixed objective:

\begin{equation}
J(\rho) = \frac{1}{N}\sum_{i = 1}^{N}\left( \rho_{i} - \rho_{i}^{\text{obs}} \right)^{2} + \lambda\frac{1}{N}\sum_{i = 1}^{N}\frac{\left( \rho_{i + 1} - \rho_{i} \right)^{2}}{\theta_{i + 1} - \ \theta_{i}}\ \
\label{eq:8}
\end{equation}

where \(\rho_{i}\) is the \(i\)th position along the interface, which
has an angular position \(\theta_{i}\), and \(\rho_{i}^{\text{obs}}\) is the
associated pixel center. \(N\) is the number of angular positions that
cover one revolution, with \(\rho_{N + 1\ } = \ \rho_{1\ }\)and
\(\theta_{N + 1\ } = \ \theta_{1\ } + \ 2\pi\). \(\lambda\) is an
arbitrary weight between the first term, which measures deviation from
the pixel center, and the second term, which penalizes larger slopes.

This is shown for the circle case in Fig.~\ref{fig:3}aii, for cases of
\(\lambda = 0\) (no smoothing), \(\lambda = 1\) (mixed smoothing) and
\(\lambda = 1000\) (smoothing-dominated). In this circular case, high
\(\lambda\) can successfully reconstruct the true interface.

Fig.~\ref{fig:3}b shows an artificially roughened interface, which is generated
by setting the radius to be that of a roughened interface made through
Fourier synthesis. In this case, while the true interface again falls
inside the center-to-center bounds, and the \(\lambda = 0\) case
introduces artificial roughness, the \(\lambda = 1000\) case removes
roughness that is present in the system. The intermediate
\(\lambda = 1\) case appears to be the most representative; while it
does not express the full detail of the true interface, it does remove
much of the artificial roughness generated through the center means,
while avoiding over-smoothing true interface roughness.

We thus acquire the \(B(r)\) from circular interfaces generated with
different values of \(\lambda\), based on deviations of the
reconstructed interface from a circle fitted to the entire interface.
Since the pixels are also no longer uniformly spaced in radial
coordinates, we instead consider the roughness in terms of angular
separation; that is:

\begin{equation}
B(r) = \left\langle \left\lbrack \rho(\theta + r) - \rho(\theta) \right\rbrack^{2} \right\rangle
\label{eq:9}
\end{equation}

\begin{figure*}[htb]
\centering

\includegraphics[width=\linewidth,height=.72\textheight,keepaspectratio]{\detokenize{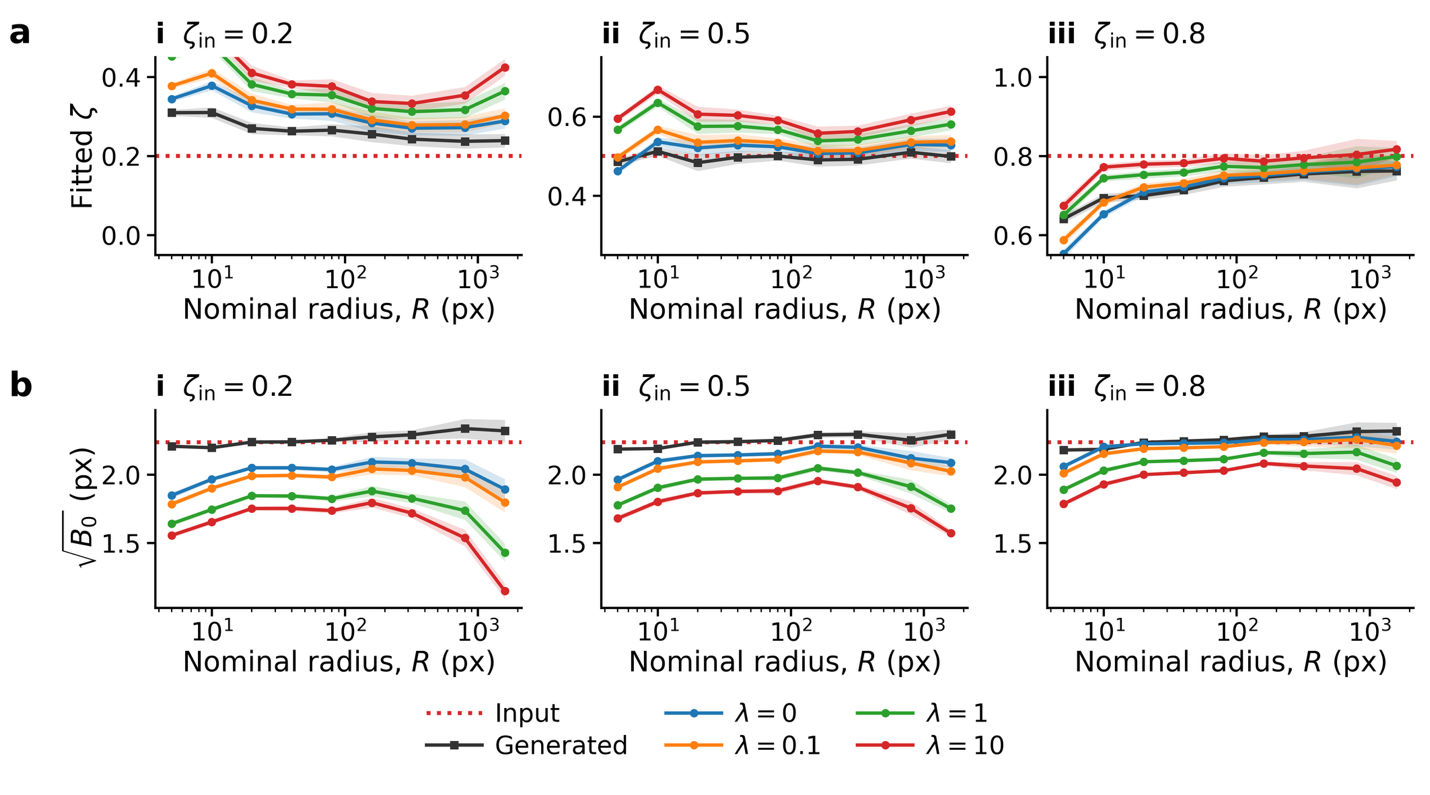}}

\caption{Recovery of circular-interface roughness at a fixed physical
reference separation \(r_{0} = 3\,\mathrm{px}\). (a) Shows the exponent
\(\zeta\) obtained from displacement-correlation fits; (b) shows the
corresponding amplitude \(\sqrt{B_{0}}\). (i--iii) correspond to
prescribed exponents \(\zeta = 0.2,\ 0.5,\ 0.8\), with
\(B_{0} = 5\,\mathrm{px}^{2}\). Results are plotted against nominal
radius \(R\). Red dotted lines indicate the prescribed values, and solid
black curves show estimates from the generated interfaces before
rasterization. The remaining curves show estimates after rasterization
and reconstruction with \(\lambda = 0,\ 0.1,\ 1,\ 10\).}

\label{fig:5}

\par\medskip

\centering

\includegraphics[width=\linewidth,height=.72\textheight,keepaspectratio]{\detokenize{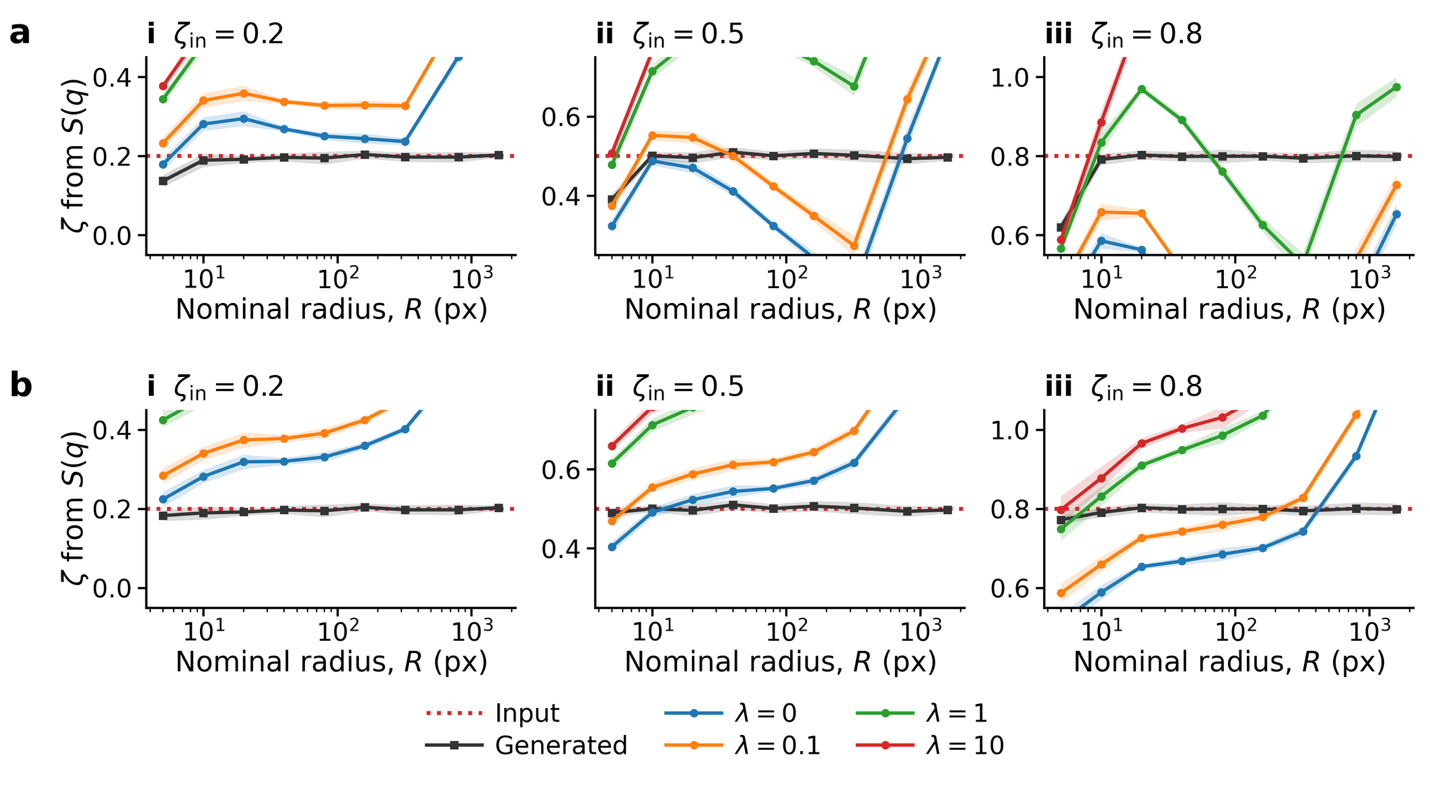}}

\caption{Spectral recovery (\(S(q)\)) of the roughness exponent from
circular interfaces. (a) Uses a fixed angular reference separation
\(r_{0} = \frac{\pi}{10}\,\mathrm{rad}\); and (b) uses a fixed physical
reference separation \(r_{0} = 3\,\mathrm{px}\). (i--iii) correspond to
prescribed exponents \(\zeta = 0.2,\ 0.5,\ 0.8\), with
\(B_{0} = 5\,\mathrm{px}^{2}\). Exponents obtained from fits to \(S(q)\)
are plotted against nominal radius \(R\). Results are plotted against
nominal radius \(R\). Red dotted lines indicate the prescribed values,
and solid black curves show estimates from the generated interfaces
before rasterization. The remaining curves show estimates after
rasterization and reconstruction with \(\lambda = 0,\ 0.1,\ 1,\ 10\).}

\label{fig:6}

\end{figure*}

\begin{figure*}[!ht]
\centering

\includegraphics[width=\linewidth,height=.72\textheight,keepaspectratio]{\detokenize{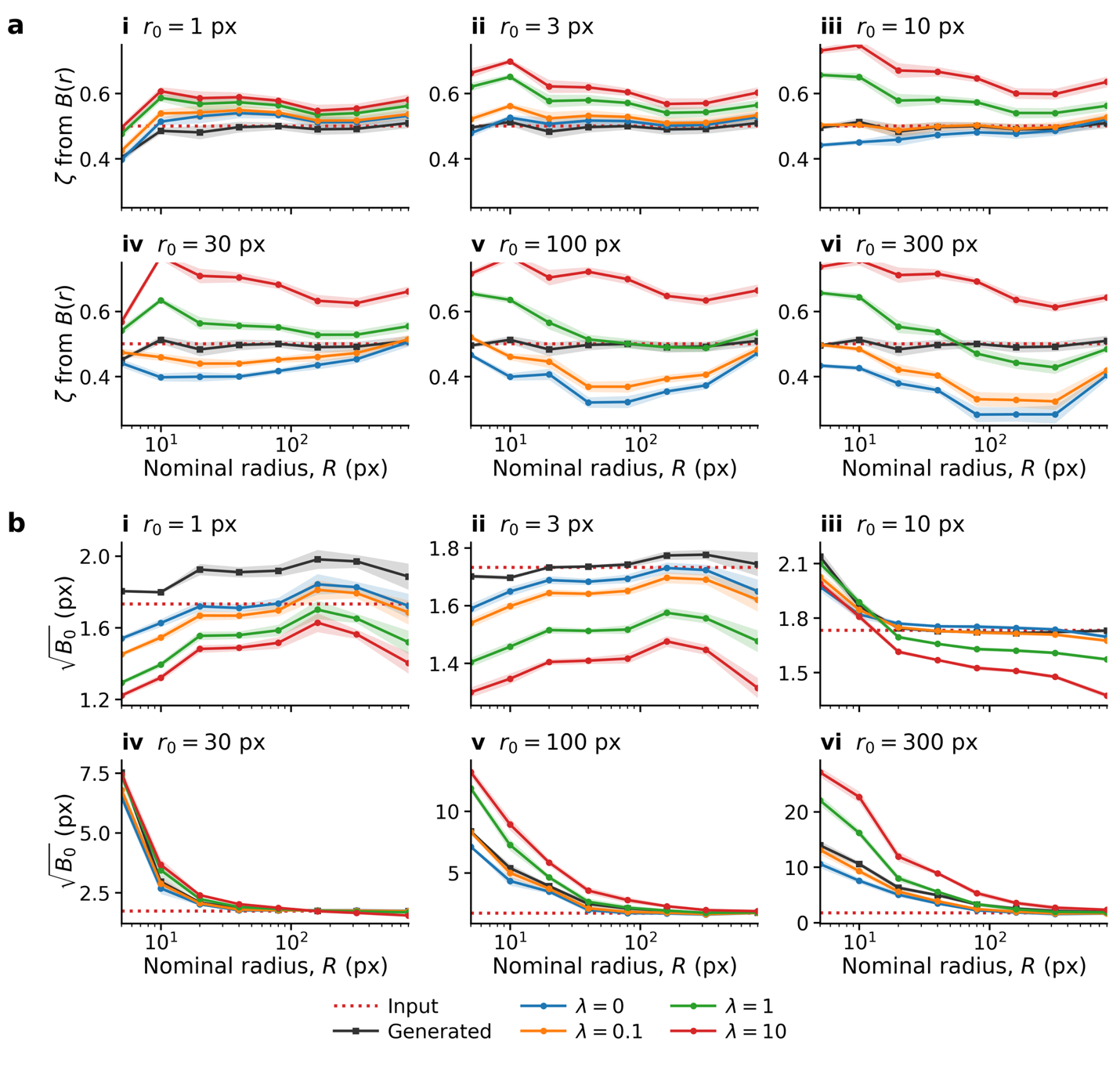}}

\caption{Dependence of roughness recovery on the physical reference
separation \(r_{0}\). (a) Shows the exponent \(\zeta\) recovered from
\(B(r)\); and (b) shows the corresponding amplitude \(\sqrt{B_{0}}\).
Within each group, \(r_{0} = 1,\ 3,\ 10,\ 30,\ 100,\ 300\,\mathrm{px}\).
The prescribed exponent and amplitude are fixed at \(\zeta = 0.5\) and
\(B_{0} = 3\,\mathrm{px}^{2}\). Red dotted lines indicate the prescribed
values, and solid black curves show estimates from the generated
interfaces before rasterization. The remaining curves show
reconstructed-interface estimates for \(\lambda = 0,\ 0.1,\ 1,\ 10\).}

\label{fig:7}

\end{figure*}

This allows avoiding the uneven spacing pixel-by-pixel. Additionally, so
as not to interpolate between pixels unnecessarily we only select for
\(r > n_{\theta}/n_{\mathrm{px}}\), where \(n_{\theta}\) and
\(n_{\mathrm{px}}\) denote the number of angle spacings and pixels
respectively. In cases where the pixel-by-pixel spacing is below the
measurement threshold, we also take the average between adjacent pixels.

\begin{figure*}[htb]
\centering

\includegraphics[width=\linewidth,height=.72\textheight,keepaspectratio]{\detokenize{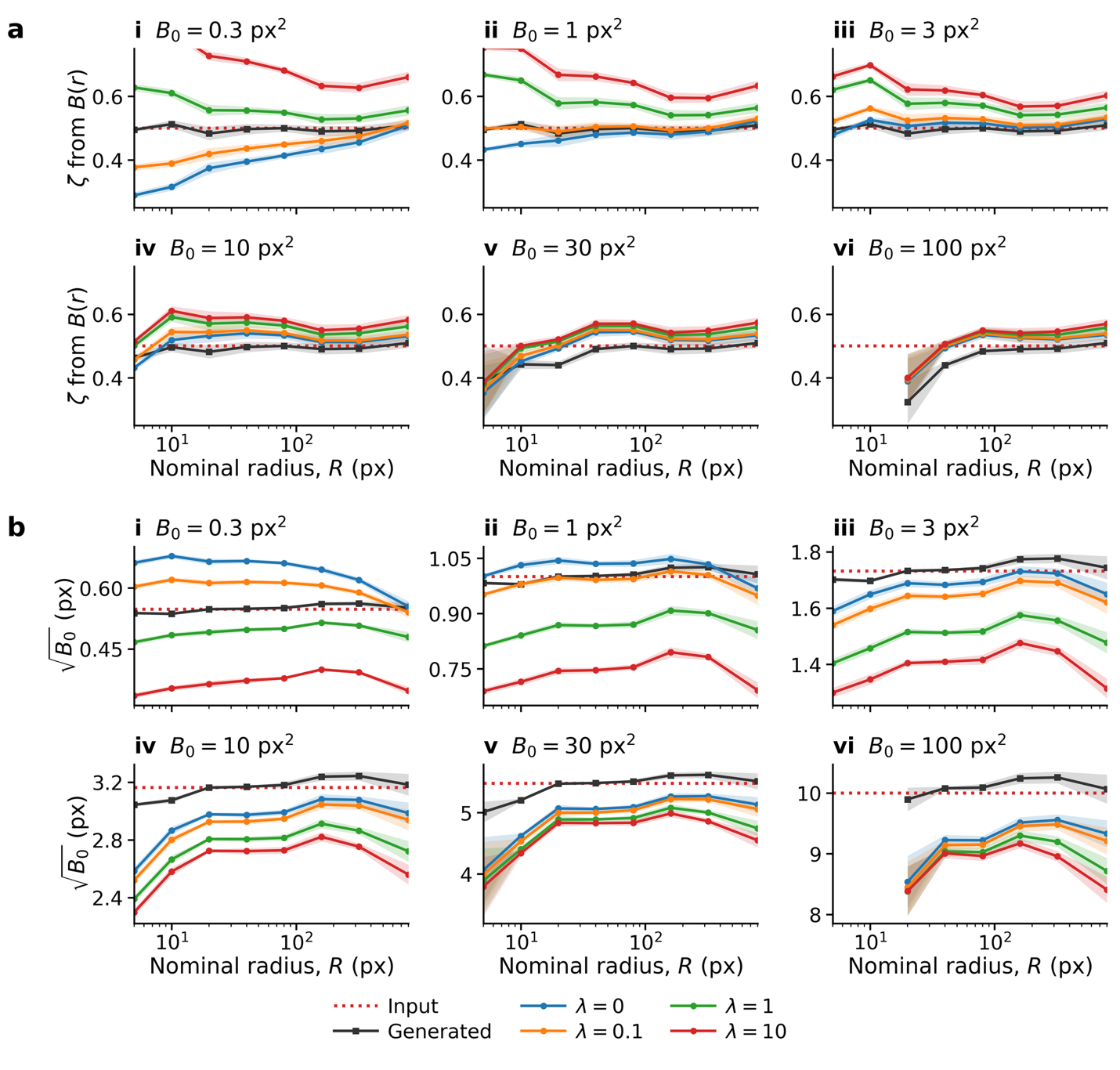}}

\caption{Dependence of roughness recovery on the prescribed amplitude
\(B_{0}\). (a) Shows the exponent \(\zeta\) recovered from \(B(r)\); and
(b) shows the corresponding amplitude \(\sqrt{B_{0}}\). Within each
group, \(B_{0} = 0.3,\ 1,\ 3,\ 10,\ 30,\ 100\,\mathrm{px}^{2}\). The
prescribed exponent and amplitude are fixed at \(\zeta = 0.5\) and
\(r_{0} = 3\,\mathrm{px}\). Red dotted lines indicate the prescribed
values, and solid black curves show estimates from the generated
interfaces before rasterization. The remaining curves show
reconstructed-interface estimates for \(\lambda = 0,\ 0.1,\ 1,\ 10\).}

\label{fig:8}

\end{figure*}

Fig.~\ref{fig:4} shows the \(\zeta\), \(B_{0}\) acquired for various values of
\(\zeta^{\text{in}}\) and smoothing term \(\lambda\) as the nominal radius
\(R\) increases. Here, \(r_{0} = \frac{\pi}{10}\,\mathrm{rad}\), and
\(B_{0} = 5\,\mathrm{px}^{2}\). These were performed such that
\(2^{12}\) samples were given per radius, and repeated 8 times to
analyze a distribution of results. Here, even before plotting the
circular interface on a raster grid, there is some difference between
\(\zeta\) and \(\zeta^{\text{in}}\), as previously shown in Fig.~\ref{fig:2}. Increasing
\(\lambda\) generally increases the measured \(\zeta\), however,
increasing agreement with \(\zeta^{\text{in}}\) at higher values. For these
cases, a finite \(\lambda\) is also necessary to reconstruct \(\zeta\);
in Fig.~\ref{fig:4}a(iii), lower values of \(\lambda\) lead \(\zeta\) to diverge
greatly from \(\zeta^{\text{in}}\).


\begin{figure*}[htb]
\centering

\includegraphics[width=\linewidth,height=.72\textheight,keepaspectratio]{\detokenize{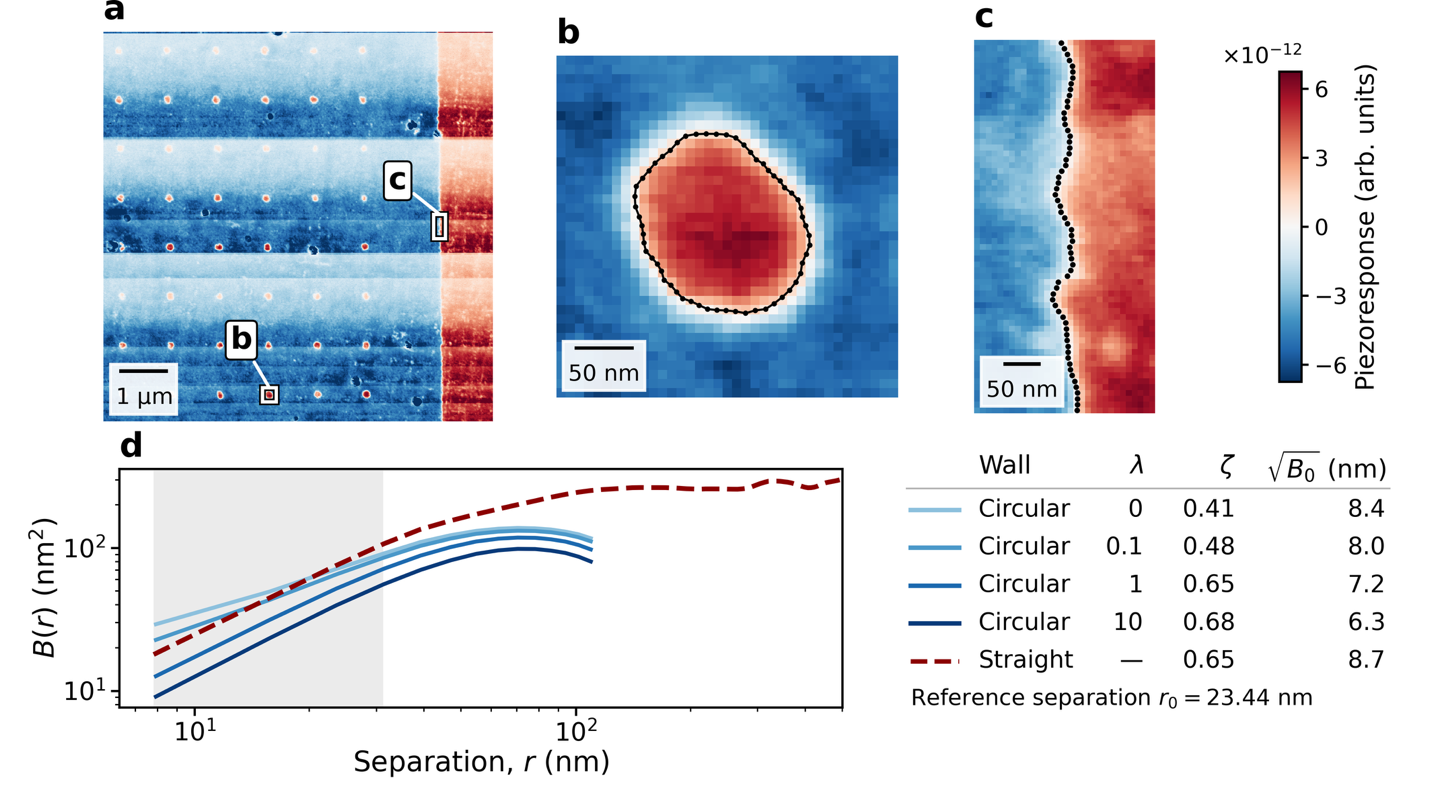}}

\caption{Roughness of probe-written ferroelectric domain walls in
BaTiO₃. (a) Signed piezoresponse map of an 8 × 8 µm² region sampled at
7.81 nm/pixel. Boxes mark the regions enlarged in (b) and (c). (b)
Quasi-circular domain with its reconstructed boundary (\(\lambda = 1\))
overlaid (c) Segment of the nominally straight wall. Each point marks
the center of a tanh fit to the piezoresponse across one scan row. (d)
Mean-square displacement \(B(r)\) versus separation \(r\) for the
ensemble of 46 quasi-circular walls (blue) and the 8 µm straight wall
(dashed red). The shaded region marks the fit window. The table gives
the roughness exponent \(\zeta\) and the rms displacement difference
\(\sqrt{B_{0}}\ \)at \(r_{0}\) = 23.44 nm.}

\label{fig:9}

\end{figure*}

In this analysis, we have considered \(r_{0}\) to be some fixed angular
separation, and as such, an arc length that increases with domain
radius. This therefore probes different physical scales as the domain
grows. In real physical phenomena, features may be restricted to a
length scale that does not grow with domain radius, such as defect
spacing in ferroelectric materials. As such, we also consider a fixed
physical separation \(r_{0}\) on a rasterized grid (where the angular
spacing decreases with longer interfaces). The results of this study,
with \(r_{0} = 3\,\mathrm{px}\), and all other variables equal are shown
in Fig.~\ref{fig:5}. For these datasets, strong agreement between the input and
measured \(\zeta\) is evident, particularly for larger radii \(R\). In
this case, however, while increasing \(\lambda\) can increase the
agreement between \(\zeta^{\text{in}}\) and \(\zeta\), extensive smoothing
causes a reduction in measured \(B_{0}\).


Note that while a fit of \(S(q)\) was more effective at identifying
\(\zeta\) than a fit of \(B(r)\) for straight interfaces, the opposite
appears true in this study of rounded interfaces. This is shown in Fig.~\ref{fig:6} for \(r_{0} = \frac{\pi}{10}\,\mathrm{rad}\) (Fig. 6a) and
\(r_{0} = 3\,\mathrm{px}\) (Fig. 6b). We attribute these differences to
inaccuracies in the rounded interface. Since many of the errors that are
introduced are sinusoidal (such as in the fit to find the center, or
pixel-by-pixel error), this can also create correlated noise in
\(a_{k}\), \(b_{k}\), in turn leading to substantial inaccuracy in
\(S(q)\) and a measured \(\zeta\).


The importance of choice of \(\lambda\) is made more prominent with
different values of \(B_{0}\) and \(r_{0}\). Fig.~\ref{fig:7} shows the evolution
of the measured \(\zeta\) with a change in radius for \(\zeta = 0.5\),
\(B_{0} = 3\,\mathrm{px}^{2}\) at different \(r_{0}\). When \(r_{0}\) is
small (near pixel resolution), \(\lambda\) does not have a strong effect
on measured \(\zeta\). However, at high \(r_{0}\), when the
pixel-to-pixel variation dominates high-frequency noise,
\(\lambda \approx 1\) minimizes this noise and produces better agreement
with the input value. Measured \(B_{0}\) tends to be highly dependent on
all variables: \(r_{0}\), \(R\), and \(\lambda\). A suitable measurement
of \(B_{0}\) thus seems to require a large dataset, so as to approach
the straight-line-interface case.


Fig.~\ref{fig:8} instead shows the evolution of the measured \(\zeta\) with a
change in radius for \(\zeta = 0.5\), \(r_{0} = 3\,\mathrm{px}\) at
different \(B_{0}\). In this case, \(\lambda \approx 1\) with a small
\(B_{0}\) (where \(\sqrt{B_{0}}\) is near or below pixel resolution) can
allow both \(\zeta\) and \(B_{0}\) to be recovered. At higher \(B_{0}\),
however, \(\lambda = 0\) produces a better result.


Finally, to demonstrate this methodology's suitability in practical
experiments, we applied this analysis to piezoresponse force microscopy
images on ferroelectric barium titanate (Fig. 9). The BaTiO\textsubscript{3}/La\textsubscript{0.7}Sr\textsubscript{0.3}MnO\textsubscript{3} heterostructure was grown by pulsed laser deposition (PLD) on a (001)-oriented SrTiO\textsubscript{3} substrate. Two types of boundaries were artificially constructed by applying bias through a conductive probe: quasi-circular domains were constructed by placing the
probe in contact with the sample, and applying a -8 V pulse for 1 s;
while nominally straight boundaries were constructed by rastering the
probe across a 10 µm × 10 µm region, and applying a -8 V voltage through
the probe at the same position along the fast-scan axis. The interface
was allowed to relax by applying a 1 V ac voltage, then an 8 µm × 8 µm
square was imaged at a resolution of 7.8125 nm/px to produce a 1024 px ×
1024 px. The interface roughness was then calculated for the 46
remaining quasi-circular domains as a single dataset (one quasi-circular
interface is shown in Fig.~\ref{fig:9}b) and an 8 µm linear segment (a segment is
shown in Fig. 9c); the final roughness parameters are described in Fig.
9d. In these experimental datasets, without any smoothing
(\(\lambda = 0\)), the scaling parameter \(\zeta\) measured from the
circular dataset is quite distinct from that extracted from the linear
domain boundary. With increased smoothing (\(\lambda\ \sim\ 1\)), the
scaling parameter is more accurately extracted. On the other hand, the
increased smoothing artificially decreases the roughness amplitude
\(\sqrt{B_{0}}\).; from 8.4 nm to 6.3 nm, when in the linear domain it
was at 8.7 nm. These results are in general agreement with Fig. 8, where
intermediate smoothing provides appropriate recovery of the scaling
parameter, but excessive smoothing artificially reduces the measured
roughness amplitude.


In summary, we have benchmarked the recovery of roughness parameters
from finite-resolution circular interfaces using both synthetic and
experimental boundaries with prescribed scaling properties. While
\(S(q)\) is the most effective method at restoring the roughness
parameters for finite straight interfaces, it is unable to restore
suitable parameters for circular interfaces within this procedure. When
these circular interfaces are large (a radius greater than
\(10^{3}\,\mathrm{px}\)), the scaling parameter \(\zeta\) can be
restored to near \(\pm 0.05\) of the prescribed exponent. However,
rasterizing the interface leads to large deviations in the scaling
parameter \(\zeta\) for particularly large reference distances \(r_{0}\)
or small roughness amplitudes \(B_{0}\). In this case, mild to moderate
smoothing by minimizing the squared gradient can improve restoration of
the roughness parameters.

These findings show that it is feasible to measure the roughness of
circular interfaces with finite resolution, but caution is required.
Smoothing parameters can improve recovery of the scaling parameter
\(\zeta\) in low-resolution datasets, but extensive repetitions and
statistics are still needed. On the other hand, excessive smoothing can
lead to an inflation of the measured \(\zeta\) and a reduction in the
measured \(B_{0}\). Analysis of circular datasets with radius
\(< 10\,\mathrm{px}\), or fitting via \(S(q)\) should be avoided with
this methodology. With these constraints, however, the roughness and
growth of circular interfaces can be more accurately studied, allowing
for more reliable interpretation of evolving circular boundaries.

\begin{acknowledgments}

This research was performed at and supported by the Center for Nanophase
Materials Sciences (CNMS), which is a US Department of Energy, Office of
Science User Facility at Oak Ridge National Laboratory.

\end{acknowledgments}

\section*{Data Availability}

The data and code presented in this work is available on
GitHub~\cite{ref22}.

\bibliography{references}

@article{ref1,
  author = {Guyonnet, Jill and Agoritsas, Elisabeth and Bustingorry, Sebastian and Giamarchi, Thierry and Paruch, Patrycja},
  title = {{Multiscaling analysis of ferroelectric domain wall roughness}},
  journal = {Physical review letters},
  year = {2012},
  volume = {109},
  number = {14},
  pages = {147601},
}

@article{ref2,
  author = {Jord{\'{a}}n, Declan and Albornoz, Lucas J and Gorchon, Jon and Lambert, Charles-Henri and Salahuddin, Sayeef and Bokor, Jeffrey and Curiale, Javier and Bustingorry, Sebastian},
  title = {{Statistically meaningful measure of domain-wall roughness in magnetic thin films}},
  journal = {Physical Review B},
  year = {2020},
  volume = {101},
  number = {18},
  pages = {184431},
}

@article{ref3,
  author = {Paruch, Patrycja and Giamarchi, Thierry and Triscone, J-M},
  title = {{Domain wall roughness in epitaxial ferroelectric PbZr 0.2 Ti 0.8 O 3 thin films}},
  journal = {Physical review letters},
  year = {2005},
  volume = {94},
  number = {19},
  pages = {197601},
}

@article{ref4,
  author = {Paruch, Patrycja and Guyonnet, Jill},
  title = {{Nanoscale studies of ferroelectric domain walls as pinned elastic interfaces}},
  journal = {Comptes Rendus. Physique},
  year = {2013},
  volume = {14},
  number = {8},
  pages = {667--684},
}

@article{ref5,
  author = {Haavig Bakke, Jan {\O}ystein and Hansen, Alex},
  title = {{Accuracy of roughness exponent measurement methods}},
  journal = {Physical Review E---Statistical, Nonlinear, and Soft Matter Physics},
  year = {2007},
  volume = {76},
  number = {3},
  pages = {031136},
}

@article{ref6,
  author = {Schmittbuhl, Jean and Vilotte, Jean-Pierre and Roux, St{\'{e}}phane},
  title = {{Reliability of self-affine measurements}},
  journal = {Physical Review E},
  year = {1995},
  volume = {51},
  number = {1},
  pages = {131},
}

@article{ref7,
  author = {Bustingorry, Sebastian and Guyonnet, Jill and Paruch, Patrycja and Agoritsas, Elisabeth},
  title = {{A numerical study of the statistics of roughness parameters for fluctuating interfaces}},
  journal = {Journal of Physics: Condensed Matter},
  year = {2021},
  volume = {33},
  number = {34},
  pages = {345001},
}

@article{ref8,
  author = {Lechenault, Fr{\'{e}}d{\'{e}}ric and Pallares, Ga{\"{e}}l and George, Matthieu and Rountree, C and Bouchaud, E and Ciccotti, Matteo},
  title = {{Effects of finite probe size on self-affine roughness measurements}},
  journal = {Physical review letters},
  year = {2010},
  volume = {104},
  number = {2},
  pages = {025502},
}

@article{ref9,
  author = {Ne{\v{c}}as, David and Valtr, Miroslav and Klapetek, Petr},
  title = {{How levelling and scan line corrections ruin roughness measurement and how to prevent it}},
  journal = {Scientific reports},
  year = {2020},
  volume = {10},
  number = {1},
  pages = {15294},
}

@article{ref10,
  author = {Fert, Albert and Reyren, Nicolas and Cros, Vincent},
  title = {{Magnetic skyrmions: advances in physics and potential applications}},
  journal = {Nature Reviews Materials},
  year = {2017},
  volume = {2},
  number = {7},
  pages = {17031},
}

@article{ref11,
  author = {Govinden, Vivasha and Prokhorenko, Sergei and Zhang, Qi and Rijal, Suyash and Nahas, Yousra and Bellaiche, Laurent and Valanoor, Nagarajan},
  title = {{Spherical ferroelectric solitons}},
  journal = {Nature materials},
  year = {2023},
  volume = {22},
  number = {5},
  pages = {553--561},
}

@article{ref12,
  author = {Catalan, Gustau and Seidel, J and Ramesh, Ramamoorthy and Scott, James F},
  title = {{Domain wall nanoelectronics}},
  journal = {Reviews of Modern Physics},
  year = {2012},
  volume = {84},
  number = {1},
  pages = {119--156},
}

@article{ref13,
  author = {Sharma, Pankaj and Moise, Theodore S and Colombo, Luigi and Seidel, Jan},
  title = {{Roadmap for ferroelectric domain wall nanoelectronics}},
  journal = {Advanced Functional Materials},
  year = {2022},
  volume = {32},
  number = {10},
  pages = {2110263},
}

@article{ref14,
  author = {Domenichini, Pablo and Pasquini, Gabriela and Capeluto, Mar{\'{\i}}a Gabriela},
  title = {{Statistical domain wall roughness analysis through correlations}},
  journal = {Physical Review B},
  year = {2023},
  volume = {108},
  number = {22},
  pages = {224434},
}

@article{ref15,
  author = {Caballero, Nirvana},
  title = {{Degradation of domains with sequential field application}},
  journal = {Journal of Statistical Mechanics: Theory and Experiment},
  year = {2021},
  volume = {2021},
  number = {10},
  pages = {103207},
}

@article{ref16,
  author = {Bulanadi, Ralph and Baxter, Jefferey and Biswas, Arpan and Funakubo, Hiroshi and Nagarajan, Valanoor and Sando, Daniel and Meier, Dennis and Schulthei{\ss}, Jan and Vasudevan, Rama and Liu, Yongtao},
  title = {{Beyond Scalar Objectives: Expert-Feedback-Driven Autonomous Experimentation for Scientific Discovery at the Nanoscale}},
  journal = {ACS nano},
  year = {2026},
  volume = {20},
  number = {41},
  pages = {27774--27783},
}

@misc{ref17,
  author = {Liu, Yongtao and Chowdhury, Jawad and Narasimha, Ganesh and Bulanadi, Ralph and Collins, Liam and Solsona, Ruben Millan and Checa, Marti and Haque, Asraful and Harris, Sumner B and Jesse, Stephen},
  title = {{AEcroscopyWave: Towards Self-Driving Characterization Platforms for Agentic AI}},
  year = {2026},
  url = {https://arxiv.org/abs/2607.22975},
  eprint = {2607.22975},
  archivePrefix = {arXiv},
}

@article{ref18,
  author = {Liu, Yongtao and Roccapriore, Kevin and Checa, Marti and Valleti, Sai Mani and Yang, Jan-Chi and Jesse, Stephen and Vasudevan, Rama K},
  title = {{AEcroscopy: a software--hardware framework empowering microscopy toward automated and autonomous experimentation}},
  journal = {Small Methods},
  year = {2024},
  volume = {8},
  number = {10},
  pages = {2301740},
}

@misc{ref19,
  author = {Chowdhury, Jawad and Harris, Sumner B and Narasimha, Ganesh and Dyck, Ondrej and Ganesh, Panchapakesan and Solsona, Ruben Millan and Checa, Marti and Collins, Liam and Bagchi, Soumendu and Vasudevan, Rama},
  title = {{Self-Driving Microscopy with Agents: Progress and Prospects}},
  year = {2026},
}

@article{ref20,
  author = {Narasimha, Ganesh and Bulanadi, Ralph and Chowdhury, Jawad and Vasudevan, Rama and Liu, Yongtao},
  title = {{Self-Driving Scanning Probe Microscopy: From Acceleration to Discovery and Manipulation}},
  journal = {Accounts of Chemical Research},
  year = {2026},
}

@article{ref21,
  author = {Makse, Hern{\'{a}}n A and Havlin, Shlomo and Schwartz, Moshe and Stanley, H Eugene},
  title = {{Method for generating long-range correlations for large systems}},
  journal = {Physical Review E},
  year = {1996},
  volume = {53},
  number = {5},
  pages = {5445},
}

@misc{ref22,
  author = {Bulanadi, Ralph},
  title = {{Finite-Circular-Interfaces}},
  year = {2026},
  howpublished = {\url{https://github.com/rbulanadi/Finite-Circular-Interfaces}},
}

\end{document}